\documentclass[aps,prl,groupedaddress, superscriptaddress,longbibliography,reprint,twocolumn]{revtex4-2}

\usepackage{amsmath}  
\usepackage{amssymb}
\usepackage{amsthm}
\usepackage{bbm}
\usepackage{graphicx}
\usepackage{color}
\usepackage{upgreek}
\usepackage[linkcolor = blue, citecolor = blue, urlcolor = blue, colorlinks = true]{hyperref}

\usepackage{amssymb}
\usepackage{bm}
\usepackage[capitalise]{cleveref}
\usepackage{textcomp}
\usepackage{hyperref}
\usepackage[usenames,dvipsnames]{xcolor}
\usepackage[normalem]{ulem}
\usepackage{wasysym}

\usepackage{floatrow}
\usepackage{wrapfig}
\usepackage{pgfgantt}

\usepackage{mathptmx}
\usepackage{bm}

\usepackage{fancyhdr}
\usepackage{blindtext}
\fancypagestyle{mypagestyle}{
\fancyhf{}
\fancyfoot[C]{\thepage}  

}
\makeatletter
\let\ps@plain\ps@mypagestyle
\makeatother
\begin{document}

\preprint{APS/123-QED}

\title{Solute-mediated colloidal vortex in a microfluidic T-junction}

\author{Haoyu Liu}
\author{Amir A. Pahlavan}%
 \altaffiliation{Corresponding author}
 \email{amir.pahlavan@yale.edu}

\affiliation{%
 Department of Mechanical Engineering and Materials Science, Yale University, New Haven, Connecticut 06511, USA\\
}%

\date{\today}

\begin{abstract}

Solute gradients next to an interface drive a diffusioosmotic flow, the origin of which lies in the intermolecular interactions between the solute and the interface. These flows on the surface of colloids introduce an effective slip velocity, driving their diffusiophoretic migration. In confined environments, we expect the interplay between diffusiophoretic migration and diffusioosmotic flows near the walls to govern the motion of colloids. These near-wall osmotic flows are, however, often considered weak and neglected. Here, using microfluidic experiments in a T-junction, numerical simulations, and theoretical modeling, we show that the interplay between osmotic and phoretic effects leads to unexpected outcomes: forming a colloidal vortex in the absence of inertial effects, and demixing and focusing of the colloids in the direction opposite to what is commonly expected from diffusiophoresis alone. We show these colloidal vortices to be persistent for a range of salt types, salt gradients, and flow rates, and establish a criterion for their emergence. Our work sheds light on how boundaries modulate the solute-mediated transport of colloids in confined environments. 

\end{abstract}

\maketitle

The discovery of colloidal migration in response to solute gradients dates back to the works of Derjaguin more than half a century ago \citep{derjaguin1961} and the later developments by Prieve, Anderson and others \citep{anderson1982, prieve1984,Ebel88, Staffeld89,Anderson1989}. While sophisticated technologies for manipulation of colloids and biomolecules in microfluidics using external electric/magnetic fields have been developed over the past two decades \citep{rousselet1994, stone2004, matsunaga2017, dreyfus2005}, using a ``pinch of salt" to guide the motion of colloids offers a tempting alternative \citep{abecassis2008,Palacci10, Palacci12,Khair13,Florea14,deseigne2014,Volk14, Kar15, Velegol16,Banerjee16,Keh16, Shin2016,Shi16,Mauger16,Schmidt16, Shin17, Shukla17,Sear17,Shin17b,Nery17,Shin17b,Ault2018, Shin18,Lee18,Raynal18,marbach2019,Gupta19,Battat19,Banerjee19,Raynal19,Ault19,Gandhi20,Gupta20,Williams20,Warren20,Shimokusu20,Singh20,shin2020,Jotkar21,Alessio21,Chu21,Shim21,Shim21b,Tan21,Doan21,shim2022,Chu22,Volk22,Alessio22,McKenzie22,Migacz22,Somasundar23,Akdeniz23,Ghosh23,Lee23,Teng23,Yang23,Sambamoorthy23,Williams24,Migacz24,Yang24} that biology seems to have benefited from as evidenced by the role of diffusiophoresis in cargo transport and phase separation within the living cells \citep{Sear19,ramm2021, Burkart22, alessio2023,Shandilya24,Hafner24,Doan24}.

Solute gradients in confined and crowded environments also drive diffusioosmotic flows next to the confining surfaces. These flows have been utilized for directional pumping and trapping of DNA and exosomes \citep{hatlo2011, lee2014,rasmussen2020}. However, they are often considered weak, leading to small changes to the diffusiophoretic migration of colloids, and therefore neglected. A few recent studies have reported on the emergence of solute-driven vortices and convection rolls near surfaces, leading to the trapping and entrainment of colloids \citep{Shin20,Chakra23,Migacz23}. Yet, how and under what circumstances diffusiophoresis and diffusioosmosis conspire to lead to these reported behaviors has remained elusive.     

Here, we use microfluidic experiments and 3D numerical simulations to demonstrate that solute-driven colloidal vortices emerge due to a subtle interplay between solutal gradients perpendicular to the walls, driving the phoretic migration of colloids toward the walls, and solutal gradients parallel to the walls, sweeping the colloids by the diffusioosmotic flows. We show the universality of these vortices for a range of salt types, salt gradients, and flow rates, and demonstrate that the vortex location, which coincides with the colloidal focusing line $x_f$ scales as $\textrm{Pe}^{-1/3}$, where $\textrm{Pe} = UH/D_s$ with $U$ as the flow velocity, $H$ as the channel height, and $D_s$ as the solute diffusivity. We finally demonstrate that the competition between diffusioosmotic mobility $\Gamma_w$ and diffusiophoretic mobility $\Gamma_p$ governs the formation of these colloidal vortices.

\begin{figure}[b]
\centering
\includegraphics[page=1, trim=10mm 10mm 5mm 15mm, clip, width=1 \columnwidth]{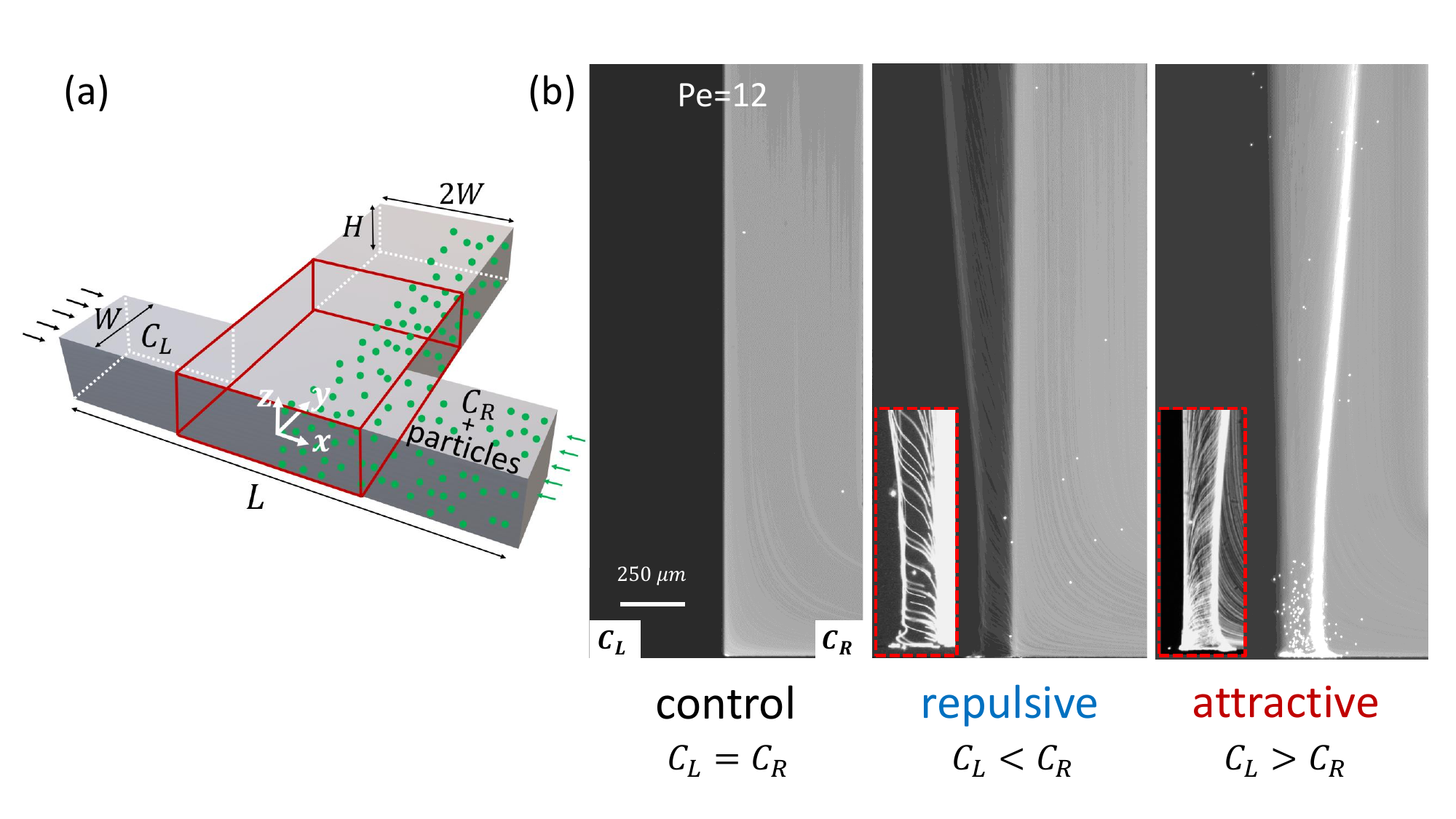}
\caption{Solute gradients lead to the emergence of a colloidal vortex and its unexpected focusing in a microfluidic T-junction. (a) Setup schematic: we inject solute concentrations $c_L$ and $c_R$ from the left and right branches, respectively. The colloidal particles (green dots) are injected from the right side. The red box demonstrates the field of view in our experiments. (b) In the absence of solute gradients, colloids simply follow the flow streamlines. Introducing a solute gradient (1mM vs 0.01mM NaCl here), however, leads to the spiral-like motion of colloids (insets), and the emergence of a focusing band in the lower solute concentration region, in contrast with the previous observations \citep{abecassis2008}.}
\label{fig1}
\end{figure}

\begin{figure*}[tbh!]
\centering
\includegraphics[page=2, trim=3mm 35mm 0mm 30mm, clip, width=1\textwidth]{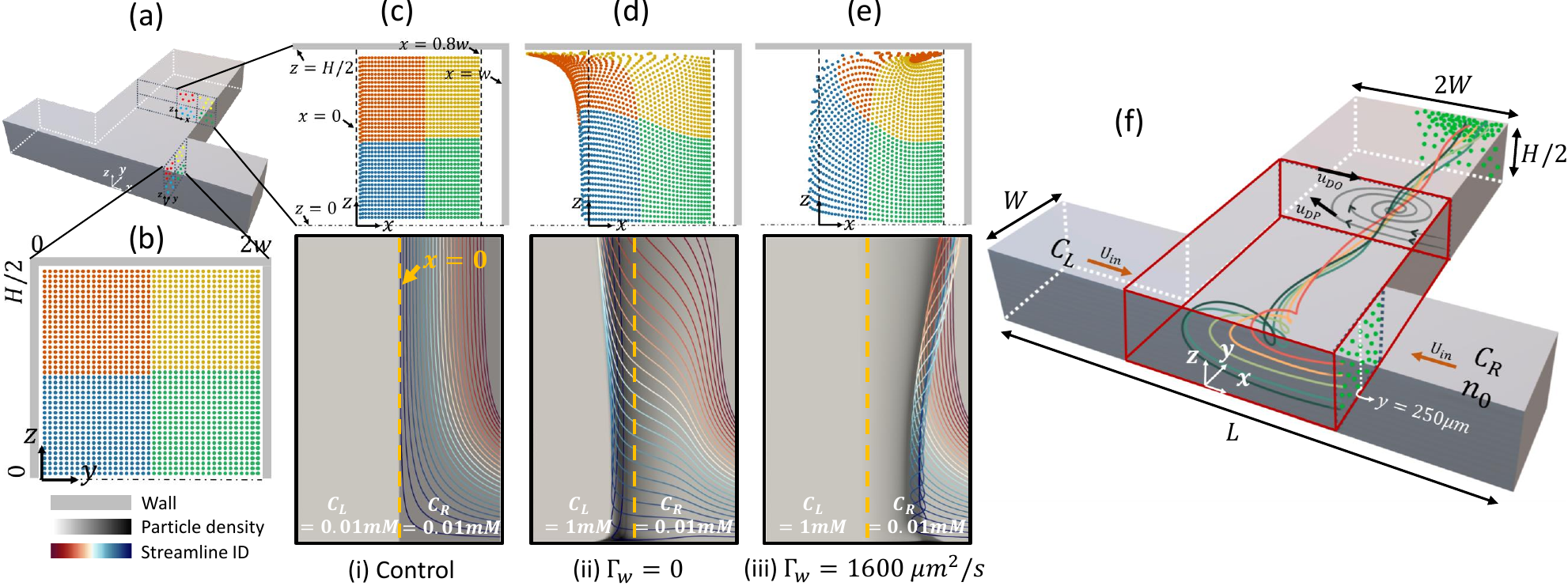}
\caption{The influence of solute gradients on the mixing of colloids and formation of spiraling trajectories. (a, b) Regularly positioned particles of four colors are released from the right branch. (c - e) Upper panel: particle distribution at a downstream cross-section. Lower panel: trajectories of 25 particles released near the wall upstream. The background gray-scale color map represents the particle density field. (c) In the control case, colloids follow the flow streamlines. (d) In the presence of diffusiophoresis alone, colloids migrate across the streamlines toward the higher solute concentration side. (e) In the presence of both diffusiophoretic and diffusioosmotic effects, spiral-like trajectories emerge. (f) Schematics of top half of the channel for the attractive case ($c_L>c_R$). The interplay between diffusiophoresis and diffusioosmosis lead to a colloidal vortex (colored solid lines) near the channel wall.}
\label{fig2}
\end{figure*}
   
We use a microfluidic T-junction setup, which is schematically shown in Fig.~\ref{fig1} (a), with height $H = 50~\mu\textrm{m}$, width $W = 500~\mu\textrm{m}$, and length
$L= 4000~\mu\textrm{m}$. The mid-channel width is set as $2W$ to keep the mean flow velocity constant. We inject solutions with salt concentrations $c_L$ and $c_R$ from the left and right sides, respectively; colloids are introduced from the right branch \citep{Liu24Sup}. In the absence of solute gradients, i.e., the control case, colloids occupy half the channel and move downstream without any noticeable diffusion (Fig.~\ref{fig1} (b)). We use particles with 1 micron diameter, with diffusivity $D_p \approx 10^{-13} \textrm{m}^2/\textrm{s}$, which should be contrasted with the solute diffusivity of $D_s \approx 10^{-9} \textrm{m}^2/\textrm{s}$. For the experiments shown in Fig.~\ref{fig1} (b), the particle Peclet number is therefore $\textrm{Pe}_p = UH/D_p \approx 10^5$, indicating the dominance of advection over diffusion for the particles.

Adding a ``pinch of salt" to either channel, however, changes the picture completely. We expect the negatively-charged colloids to migrate toward the higher salt concentration for the salts we use (LiCl, NaCl, KCl) \citep{abecassis2008}. Surprisingly, however, we observe the focusing of colloids in the lower salt concentration region, i.e., when $c_L > c_R$, colloids move to the right side, and when $c_L < c_R$, a focusing line emerges on the left side (Fig.~\ref{fig1} (b)), in contrast to what is expected \citep{abecassis2008}.

To gain insight into this unexpected focusing, we conducted 3D numerical simulations of the flow, solute, and particle transport. The flow field is governed by the Stokes equations:
\begin{align} 
0 &= -\nabla p + \mu \nabla^2 \mathbf{u},\\
0 &= \nabla \cdot \mathbf{u},
\end{align}
where $\mathbf{u}$ is the flow velocity, $p$ represents the fluid pressure, and $\mu$ is the fluid viscosity. The solute $c$ and colloid $n$ fields are governed by advection-diffusion equations:
\begin{align} 
\frac{\partial c}{\partial t} &= -\nabla \cdot\left(\mathbf{u} c\right) + D_s \nabla^2 c,\\
\frac{\partial n}{\partial t} &= -\nabla \cdot\left(\left(\mathbf{u} + \mathbf{u}_{dp}\right) n\right) + D_p \nabla^2 n,
\end{align}
where the colloid diffusiophoretic velocity is $\mathbf{u}_{dp} = \Gamma_p \nabla \ln c$ with $\Gamma_p$ as the diffusiophoretic mobility \citep{prieve1984,Anderson1989,marbach2019,Liu24Sup}. The flow velocity on the walls of the channel is determined by the diffusioosmotic slip velocity $\mathbf{u}_{do} = - \Gamma_w \left( \nabla \ln c - \left(\mathbf{e}_n \cdot \nabla \ln c \right) \mathbf{e}_n \right)$, where $\mathbf{e}_n$ is the unit normal vector on the boundary. We solve these equations using the open-source software OpenFoam \citep{weller1998,Liu24Sup}. 

\begin{figure*}[tbh!]
\centering
\includegraphics[page=3, trim=10mm 55mm 5mm 35mm, clip, width=\linewidth]{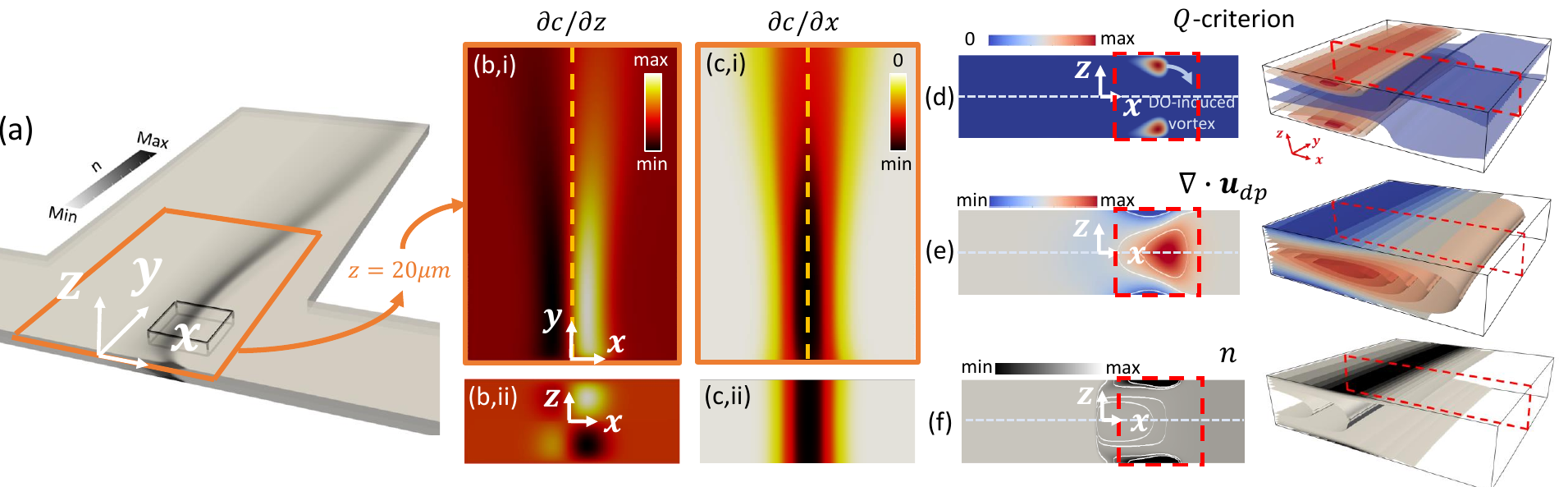}
\caption{The differential solute dispersion leads to the formation of colloidal vortices and their focusing. (a) The colloid density field, showing focusing near the walls; the orange rectangle shows the x-y plane 5 $\mu\textrm{m}$ away from the wall; the black box is the field of view corresponding to the panels (d-f), where isosurfaces are shown. The solute concentrations are $c_L=1$mM and $c_R=0.01$mM. (b,c) The presence of shear near the wall and its absence in the mid-plane leads to the differential solute dispersion, generating solute gradients both in-plane and out-of-plane. (d) The solute gradients near the wall drive a diffusioosmotic flow and create a vortex, which can be identified using the Q criterion. (e) The focusing location coincides with the region of negative divergence of diffusiophoretic velocity, leading to an effective compressibility of the colloid field, and acting as a sink for the particles. (f) The interplay between the diffusioosmotic vortex and the effective compressibility, lead to the colloidal accumulation near the wall on the lower solute concentration side.} 
\label{fig3}
\end{figure*}

In a Lagrangian frame, the colloidal transport can be described as $d\mathbf{x}_p/dt = \mathbf{u} + \mathbf{u}_{dp}$, where $\mathbf{x}_p(t)$ represents the particle location at time $t$. Following the trajectories of particles obtained from the simulations, we gain insight into the focusing mechanism (Fig.~\ref{fig2}). In the control case, where solute gradients are absent, colloids simply follow the flow streamlines as evident from the absence of mixing of the colored particles introduced upstream in the four quadrants (Fig.~\ref{fig2} (a, b)), which remain in their corresponding quadrants as they travel along the channel (Fig.~\ref{fig2} (c)). In this case, there is no colloidal motion in the z direction, perpendicular to the fluid flow.

In the presence of salt gradients, both diffusiophoretic migration of colloids and diffusioosmotic flows emerge. To disentangle these two effects, we first turn-off the osmotic flows near the walls, setting $\Gamma_w=0$. In this case, colloids deviate from the flow streamlines, and move toward the higher salt concentration side due to diffusiophoresis (Fig.~\ref{fig2} (d)). This behavior is consistent with the observations of \citet{abecassis2008}; however, opposite of what we observed in our experiments (Fig.~\ref{fig1}). We therefore conclude that diffusiophoresis alone cannot explain our observations.

Accounting for the diffusioosmotic flows near the channel walls changes the particle trajectories drastically (Fig.~\ref{fig2} (e)). The colloids move out-of-plane in the $z$ direction, and get dragged along the walls due to the diffusioosmotic flows, leading to their spiraling motion, and focusing on the right side of the channel, opposite of what diffusiophoresis alone does (Fig.~\ref{fig2} (f)). The presence of diffusioosmosis, therefore, does not simply lead to a weak quantitative drift in the colloidal trajectories, but leads to significant \textit{qualitative} changes, and an effective demixing of the colloidal field that certainly cannot be described using a diffusive framework.

The out-of-plane motion of colloids is driven by the solute gradients in the $z$ direction. The presence of shear near the wall and its absence in the mid-plane leads to a differential solute diffusion and is the reason for the emergence of solute gradients in the $z$ direction, driving the motion of colloids both within and perpendicular to the x-y plane  (Fig.~\ref{fig3} (a-c)). In the mid-plane of the channel, solute diffusion competes with advection along the channel ($U \partial c/\partial y \approx D_s \partial^2 c / \partial x^2$), leading to $\delta_s/H \sim (y/H)^{1/2} \textrm{Pe}^{-1/2}$ scaling for the solute diffusion width $\delta_s$. Near the walls, however, the presence of shear fow $\dot\gamma z$ changes this scaling ($\dot \gamma z \partial c / \partial y \approx D_s (\partial^2 c / \partial x^2 + \partial^2 c / \partial z^2)$), leading to $\delta_s/H \sim (y/H)^{1/3} \textrm{Pe}^{-1/3}$ scaling for the diffusion width \citep{Ismagilov2000, salmon2007,Liu24Sup}.  

    \begin{figure*}[tbh!]
    \centering
    \includegraphics[page=4, trim=0mm 38mm 0mm 35mm, clip, width=\linewidth]{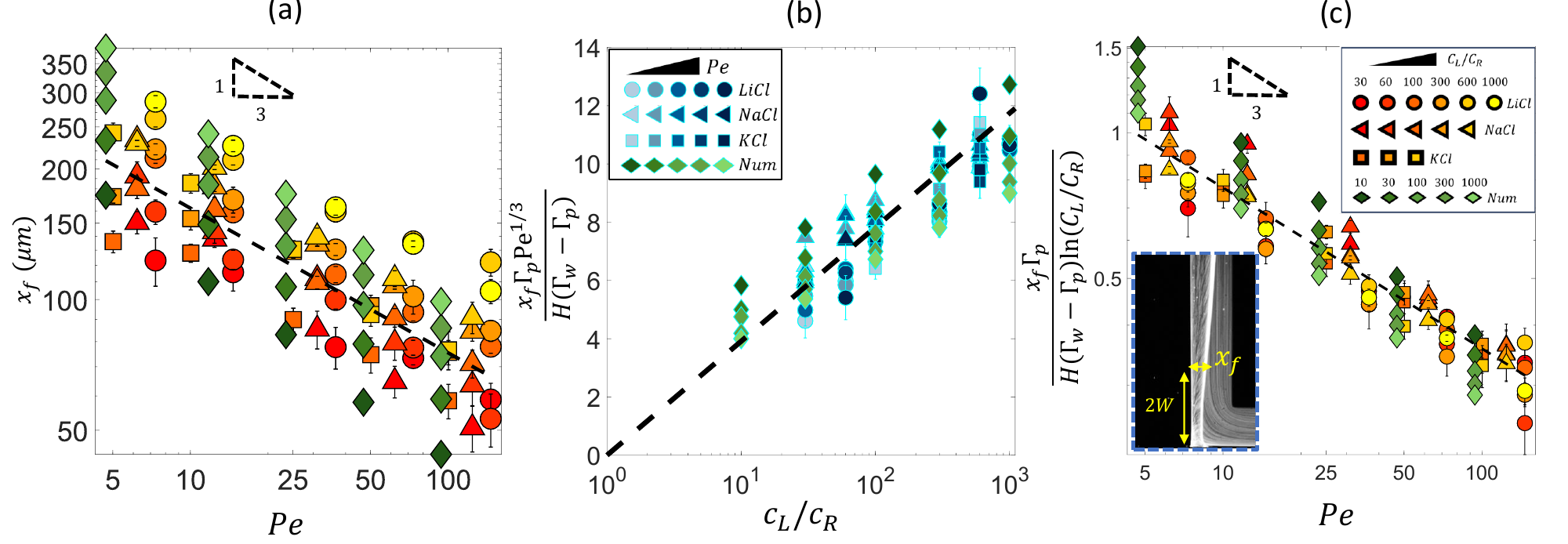}
    \caption{The focusing location is determined by the competition between diffusiophoresis, diffusioosmosis, and flow. (a) $x_f$ against Pe. Warm colored symbols represent all the experimental results for different salt solutions (circles: LiCl; triangles: NaCl; squares: KCl) with a range of concentration $c_L$ (from red: $c_L=0.3\text{mM}$ to yellow: $c_L=10\text{mM}$) and $c_R=0.01\text{mM}$. Green diamonds represent the numerical results under fixed $\Gamma_w=1600{\mu}\text{m}^2/s$, $\Gamma_p=800{\mu}\text{m}^2/s$. (b) Normalized $x_f$ by the characteristic length scale $H\text{Pe}^{-1/3}({\Gamma_w-\Gamma_p})/{\Gamma_p}$. The plot is in semi-log scale, showing that $x_f$ is proportional to $\ln (c_L/c_R)$. (c) Normalized $x_f$ by the characteristic length scale $H\ln (c_L/c_R)({\Gamma_w-\Gamma_p})/{\Gamma_p}$, showing a scaling of $\text{Pe}^{-1/3}$.}
    \label{fig4}
    \end{figure*}

Once near the walls, the colloids experience the competing effects of diffusiophoresis and diffusioosmosis. These osmotic flows create a fluid vortex, which is responsible for the spiral-like motion of the colloids \citep{Liu24Sup}. To identify these osmotic near-wall vortices, we rely on the second invariant of the velocity gradient tensor, $\mathbf{A} \equiv \nabla \mathbf{u}$, known as the Q criterion \citep{hunt1988}. The eigenvalues of the velocity gradient tensor satisfy the characteristic equation $\lambda^3 + P \lambda^2 + Q \lambda + R = 0$, where $P$, $Q$, and $R$ are the invariants \citep{jeong1995}. These invariants can be written in terms of the symmetric rate-of-strain tensor $\mathbf{S} = (\nabla \mathbf{u} + \nabla \mathbf{u}^T)/2$ and antisymmetric vorticity tensor  $\bm{\Omega} = (\nabla \mathbf{u} - \nabla \mathbf{u}^T) /2$, where $P = -\textrm{tr}(\mathbf{A}) = 0$ for incompressible flows, $R = - \textrm{det}(\mathbf{A})$, and
\begin{equation}
Q = (\| \bm{\Omega} \|^2 - \| \mathbf{S} \|^2) / 2,
\end{equation}
with $Q>0$ identifying a vortex as the region, where vorticity dominates over the rate of strain \citep{Liu24Sup} (Fig.~\ref{fig3} (d)). The colloid focusing can therefore be understood as the interplay between the diffusiophoretic migration of colloids away from the mid-plane toward the walls and entrainment by the diffusioosmotic vortex near the walls.

The focusing of colloids near the walls creates a local maximum in the colloid density field, where $\nabla n \approx 0$. The colloid evolution equation near this local maximum can therefore be simplified to
\begin{equation}
 \frac{\partial n}{\partial t} \approx \underbrace{- n \nabla \cdot \mathbf{u}_{dp}}_{\text{compressibility}} + \underbrace{D_p \nabla^2 n}_{\text{diffusion}}, 
 \end{equation}
which balances the colloid accumulation due to the effective compressibility $\nabla \cdot \mathbf{u}_{dp} < 0$ with the colloid diffusion away from the peak, where $ \nabla^2 n < 0$ (Fig.~\ref{fig3} (e,f)). This effective compressibility of the colloid density field creates an effective particle source near the mid-plane and an effective particle sink near the walls \citep{Singh20}.

The competition between diffusioosmosis and diffusiophoresis modulates the focusing location of colloids near the wall. At steady state and neglecting particle diffusion, this balance can be written as $\nabla \cdot \left( (\mathbf{u} + \mathbf{u}_{dp}) n \right) = 0$, which further simplifies to
\begin{equation}
\underbrace{\left(\Gamma_w - \Gamma_p\right) \frac{\partial \ln{c}}{\partial x}}_{\text{osmosis vs phoresis}}  \left(\frac{\partial \ln{n}}{\partial x}\right) \approx \underbrace{\Gamma_p  \frac{\partial^2 \ln{c}}{\partial x^2}}_{\text{compressibility}},
 \end{equation}
leading to the following scaling for the particle focusing location
 \begin{equation}
x_f/H \sim \left(\frac{\Gamma_w - \Gamma_p}{\Gamma_p} \right) \underbrace{\left(\frac{y}{H} \right)^{1/3} \textrm{Pe}^{-1/3}}_{\delta_s/H},
 \end{equation}
 which suggests that the solute diffusion width near the walls sets the length scale for the colloidal focusing (Fig.~\ref{fig4} (a)). The dependence of the focusing location on the strength of solute gradients can be inferred from the evolution of colloids in the Lagrangian frame near the wall, which can be written as $d x_p/dt = (\Gamma_w - \Gamma_p) \partial \ln c / \partial x$, suggesting $x_f \sim \ln (c_L/c_R)$. This scaling argument for the colloid focusing location is supported by our experimental observations and numerical simulations (Fig.~\ref{fig4} (b-c)), where $x_f$ is defined as the distance between the central line $x=0$ and the focusing band at location $y=2W$. In the experiments, $x_f$ is obtained by averaging a series of steady-state snapshots of the colloid field, and the error bars represent the standard deviation of measurements \citep{Liu24Sup}. We further demonstrate in the Supplementary Materials that this scaling is consistent with self-similarity arguments for the solute and colloid fields \citep{Liu24Sup}. 

\begin{figure}[b!]
\centering
\includegraphics[page=5, trim=10mm 10mm 2mm 10mm, clip, width=1 \columnwidth]{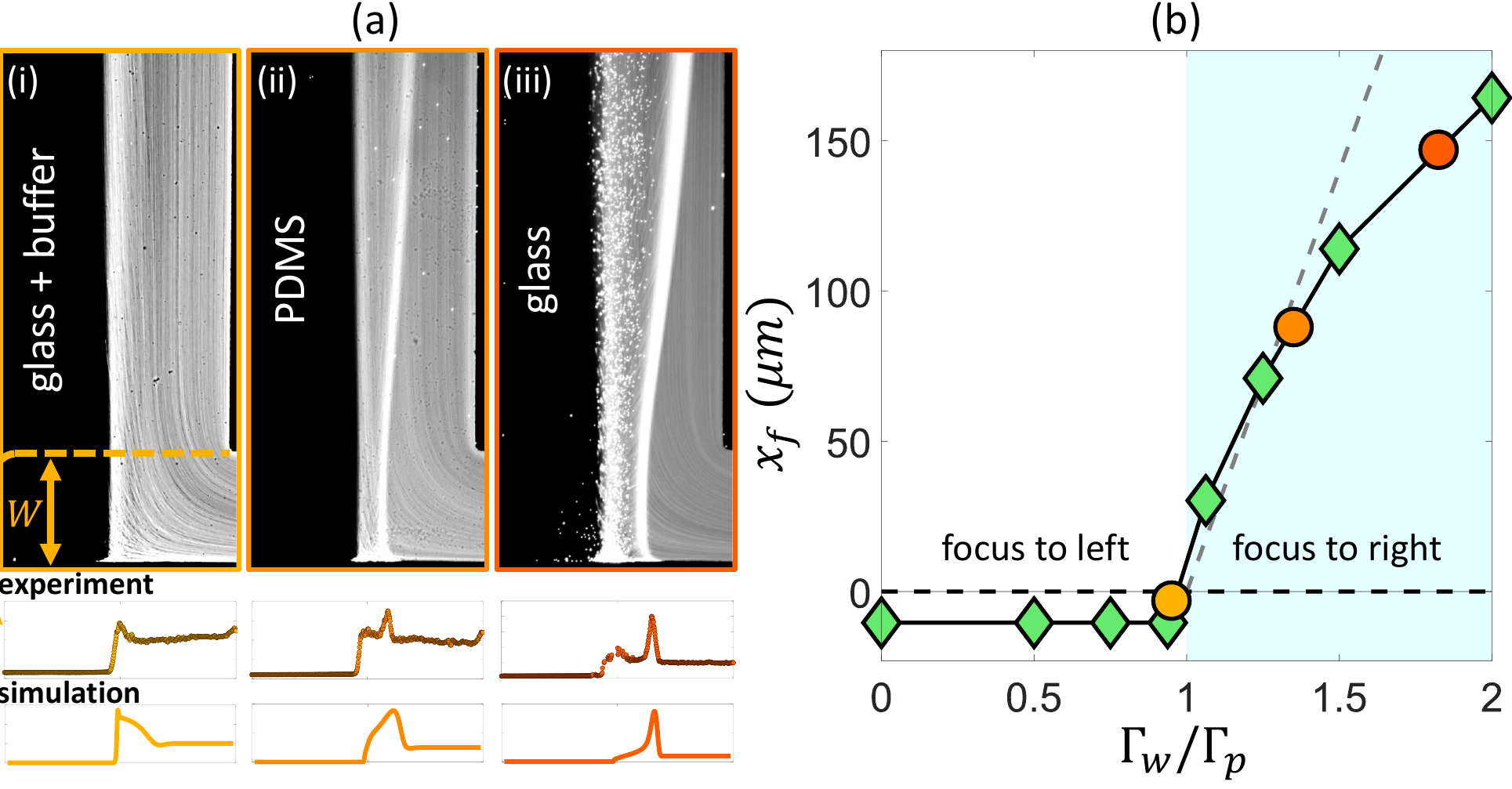}
\caption{The colloidal focusing location can be modulated by varying the ratio of diffusioosmotic and diffusiophoretic mobilities ($\Xi \equiv \Gamma_w/\Gamma_p$). (a) The diffusioosmotic mobility can be changed by using different materials: i) glass + Tris-HCl buffer with pH of 7.5 on both sides, ii) PDMS-coated glass, and iii) glass. Here, $c_L=10\text{mM}$, $c_R=0.01\text{mM}$, and $\text{Pe}\approx 12$. We observe no focusing to the right in (i), and weaker focusing in (ii) compared to (iii). The insets show the 1D cross-sectional particle intensities from the experiments and simulations. (b) We can infer the diffusioosmotic mobility of the surfaces in our experiments (circles) by comparing the focusing locations with those from the simulations with different $\Gamma_w$ values (diamonds), where $\Gamma_p=800\mu\text{m}^2/s$ is kept constant. }
\label{fig5}
\end{figure}

Our scaling argument suggests that the focusing side is determined by the ratio $\Xi \equiv \Gamma_w/\Gamma_p$: when osmotic mobility is stronger than the phoretic mobility ($\Xi>1$), focusing occurs on the lower concentration side as we have shown. However, when phoretic mobility dominates ($\Xi<1$), we expect the focusing on the higher concentration side. For a colloid with a given $\Gamma_p$ value, we can modulate the diffusioosmotic mobility of the surfaces by using different materials (glass versus PDMS) or by adding a buffer solution \citep{Liu24Sup}. We observe that adding the buffer solution completely eliminates the diffusioosmotic flows, explaining why the early work of \citet{abecassis2008} had not observed these flows (Fig.~\ref{fig5} (a,i)). On a PDMS-coated channel, we do observe focusing to the right, but weaker than that of the glass (Fig.~\ref{fig5} (a,ii-iii)). While we cannot directly measure the diffusioosmotic mobility of these surfaces, we can infer them from the comparison of the colloid focusing location in the experiments and simulations (Fig.~\ref{fig5} (b)). Our simulations show that for $\Xi<1$, the focusing location becomes insensitive to the value of diffusioosmotic mobility. However, for $\Xi>1$, the focusing location keeps moving further to the right; this increase is initially almost linear, before plateauing to an upper limit, which is determined by the solute diffusion width \citep{Liu24Sup}. We note here that we have used a constant zeta potential in all our simulations so far; considering variable zeta potential models leads to quantitative changes in our simulation results as discussed in the Supplementary Materials \citep{Liu24Sup}.

Our observations suggest that the subtle interplay between diffusiophoresis and diffusioosmosis can be utilized to infer the diffusioosmotic mobility of the surfaces, or separate particles based on their zeta potential or size. Beyond microfluidic applications for particle sorting and separation, our findings point to the importance of accounting for the often-ignored diffusioosmotic flows in the solute-mediated transport of colloids in complex environments, from contaminant removal in porous materials to cargo delivery in living cells \citep{Park21,Jotkar24,ramm2021, Burkart22,Shandilya24}. We further expect these osmotic flows to play role in modulating reactions in subsurface flows \citep{Kar16,Plumper17}, phase separation in living cells \citep{Hafner24,Doan24}, and ionic transport in porous electrodes \citep{Newman75,Biesheuvel11,Ferguson12,Mirzadeh14,Alizadeh17,Smith17,Maggiolo20,Henrique24}.

\bibliography{main}

\end{document}


\title{Supplemental Material: \\Solute-mediated colloidal vortex in a microfluidic T-junction}

\author{Haoyu Liu}
\author{Amir A. Pahlavan}%
\altaffiliation{Corresponding author}
\email{amir.pahlavan@yale.edu}

\affiliation{%
Department of Mechanical Engineering and Materials Science, Yale University, New Haven, Connecticut 06511, USA\\
}%

\date{\today}

\maketitle

\renewcommand{\thefigure}{S\arabic{figure}}

\tableofcontents

\newpage
\subsection{1. The experimental setup}

    \begin{figure}[ht!]
    \centering
    \includegraphics[page=1, trim=60mm 40mm 65mm 30mm, clip, width=0.8\columnwidth]{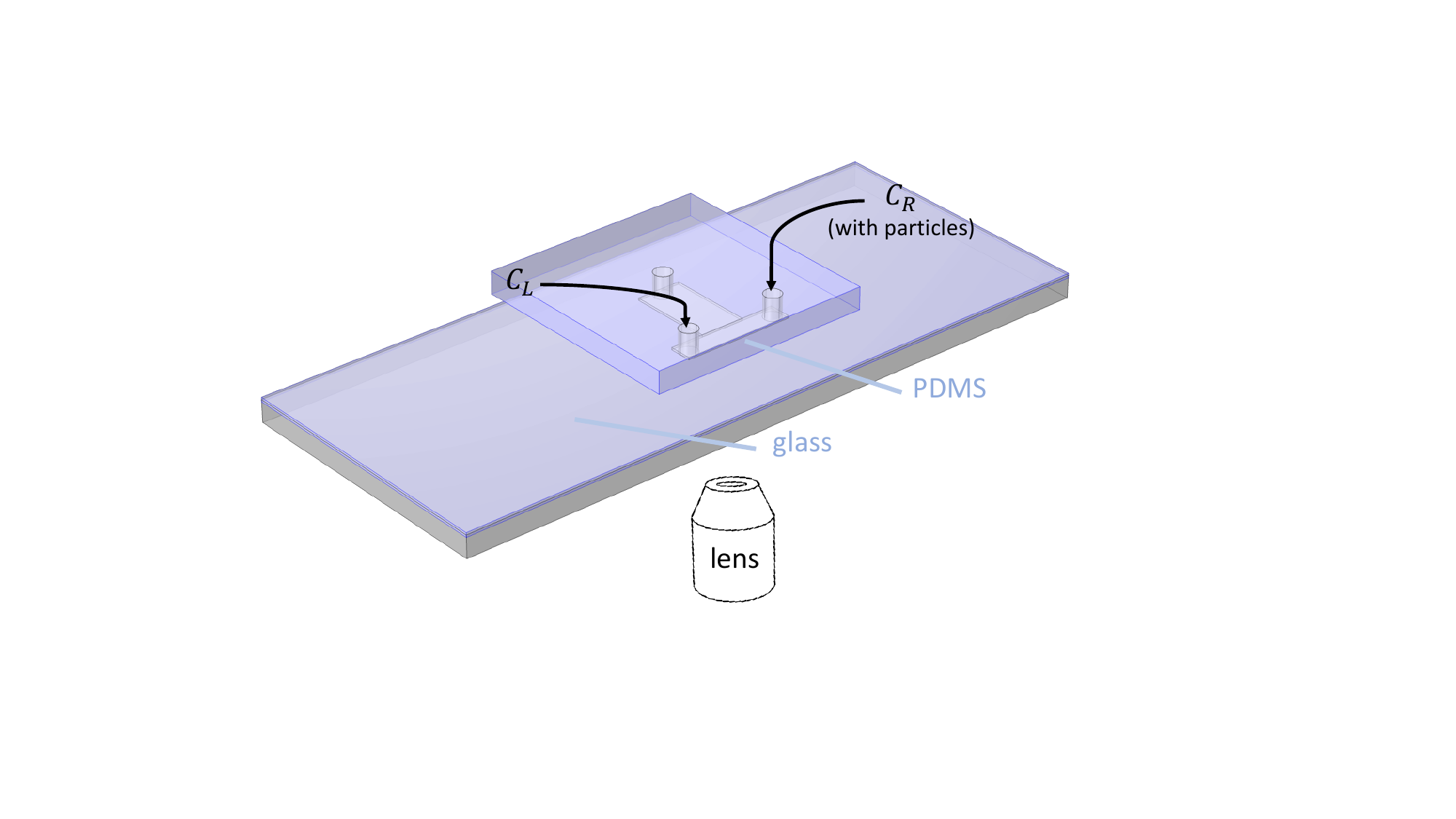}
    \caption{Schematics of the microfluidic T-junction setup.}
    \label{figs1}
    \end{figure}

Experiments are performed on microfluidic chips (Fig.~\ref{figs1}) fabricated using the standard soft lithography techniques \citep{Whitesides1998soft}. We bond the Polydimethylsiloxane (PDMS) channels either i) directly on glass, or ii) on a glass coated with a thin layer of PDMS. Glass surfaces have higher zeta potential than PDMS \citep{kirby2004zeta1, kirby2004zeta2}, leading to a stronger diffusioosmotic flow on the surface. Therefore our observations of the colloidal vortex mainly focus on the glass side.

Flow in the two sub-channels are controlled by two independent syringe pumps (Harvard Apparatus), from which we create an oscillation-free flow rate from 0.005 to 0.1${\mu}L/s$, corresponding to salt Peclet numbers $\approx 5-100$ in our experiments. Imaging was performed by an inverted fluoresence microscope (Nikon ECLIPSE Ti2). Because of the steady-state nature of the problem, images of long exposure time (2-10 seconds) are taken to show the particle trajectories clearly. Images are then analyzed using the open-source image processing software ImageJ. To obtain the colloidal focusing position $x_f$ (main text Fig.4), we average 250 images and find the peak location of the intensity profile. Error bars represent the standard deviation.

The following table shows the source of materials used in our experiments: 

    \begin{table}[ht!]
    \centering
    \begin{tabular}{c|c}
    \hline
    Chemicals & Source \\
    \hline
    Polydimethylsiloxane (PDMS) &  Dow Corning \\
    Polystyrene latex particles & Thermo Fisher Scientific \\
    Salts (LiCl, NaCl, KCl) & Sigma Aldrich \\
    DI water & Milli-Q system \\
    \hline
    \end{tabular}
    \caption{Materials source}
    \label{Materials}
    \end{table}

\subsection{2. The calibration of particle density with the light intensity}

\begin{figure}[h!]
    \centering
    \includegraphics[page=2, trim=55mm 40mm 65mm 35mm, clip, width=0.8\columnwidth]{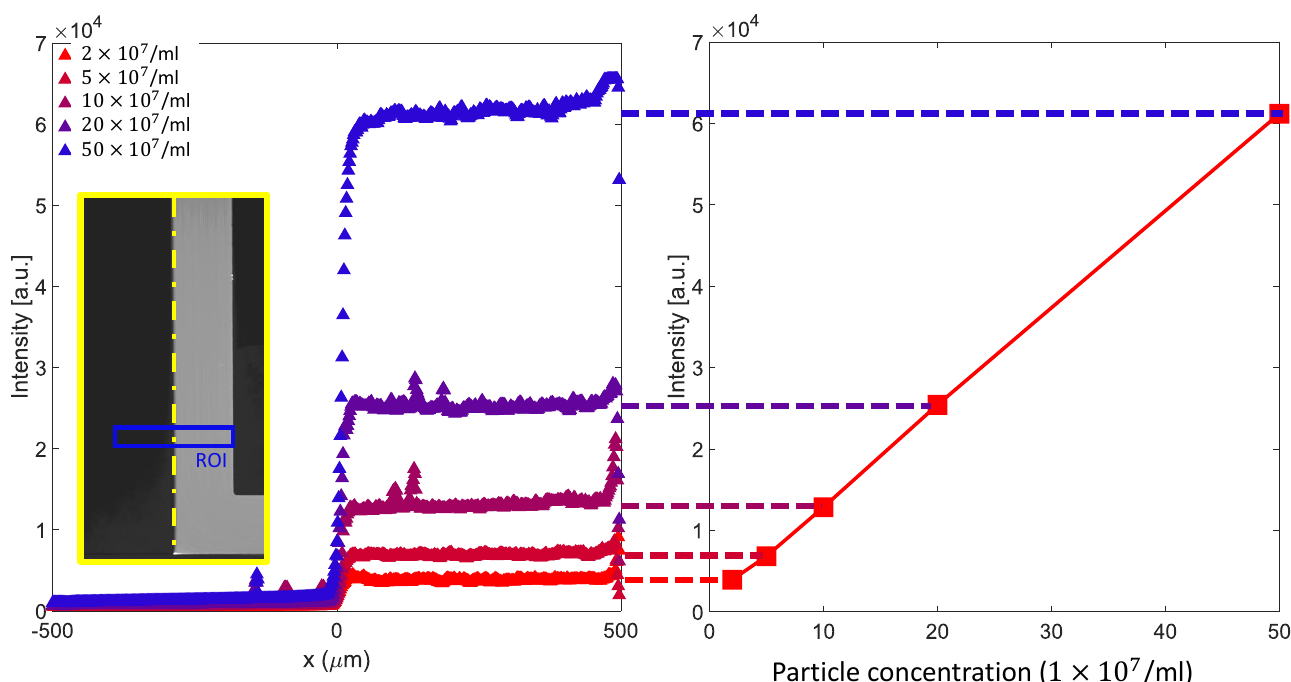}
    \caption{The calibration of particle density with the light intensity. Left: intensity profiles of different particle density (inset: experimental image, with ROI shown by the blue box). Right: light intensity versus particle concentration.}
    \label{figs6}
    \end{figure}

Our experimental images represent the light intensity emitted from the colloids within the field of view. To understand how this light intensity correlates with the particle density, we performed a series of calibration experiments. We tune the initial particle density by adding different amount of undiluted colloidal suspension into DI water, keeping the total volume equal to 1 ml. In Fig.~\ref{figs6}, we obtained the average particle intensity over 60 images, with particle number density 2, 5, 10, 20, 50 ($10^7/ml$). Left panel shows the intensity profile at $y=2W$. Right panel shows a linear relation between concentration and the intensity, within this density range. We therefore expect the light intensity in our experiments to be linearly proportional to the particle density.

\subsection{3. Numerical simulations}

All of the numerical simulations were performed by finite-volume method using the open-source software OpenFOAM \citep{weller1998}. We created an in-house steady-state solver simpleOsmoFOAM, combining the simpleFOAM and scalarTransportFOAM, which are used for solving steady flow field and advection-diffusion transport of solute, respectively. For implementing the slip diffusioosmotic boundary condition coupled with the solute transport, we utilized groovyBC in swak4FOAM to add a sliding-wall condition at the wall \citep{AultJFM2018,Ault19,battat2019particle}. In the step of solving fluid flow and solute transport, the sliding-wall velocity is updated to be $-\Gamma_w{\nabla}\ln c$ for every iteration. The flow field and solute distribution are solved together, and they finally reach a steady state, based on which we can then either solve for the continuous particle density or analyze the particle trajectories.

\begin{figure}[h!]
    \centering
    \includegraphics[page=3, trim=18mm 35mm 25mm 35mm, clip, width=0.8\columnwidth]{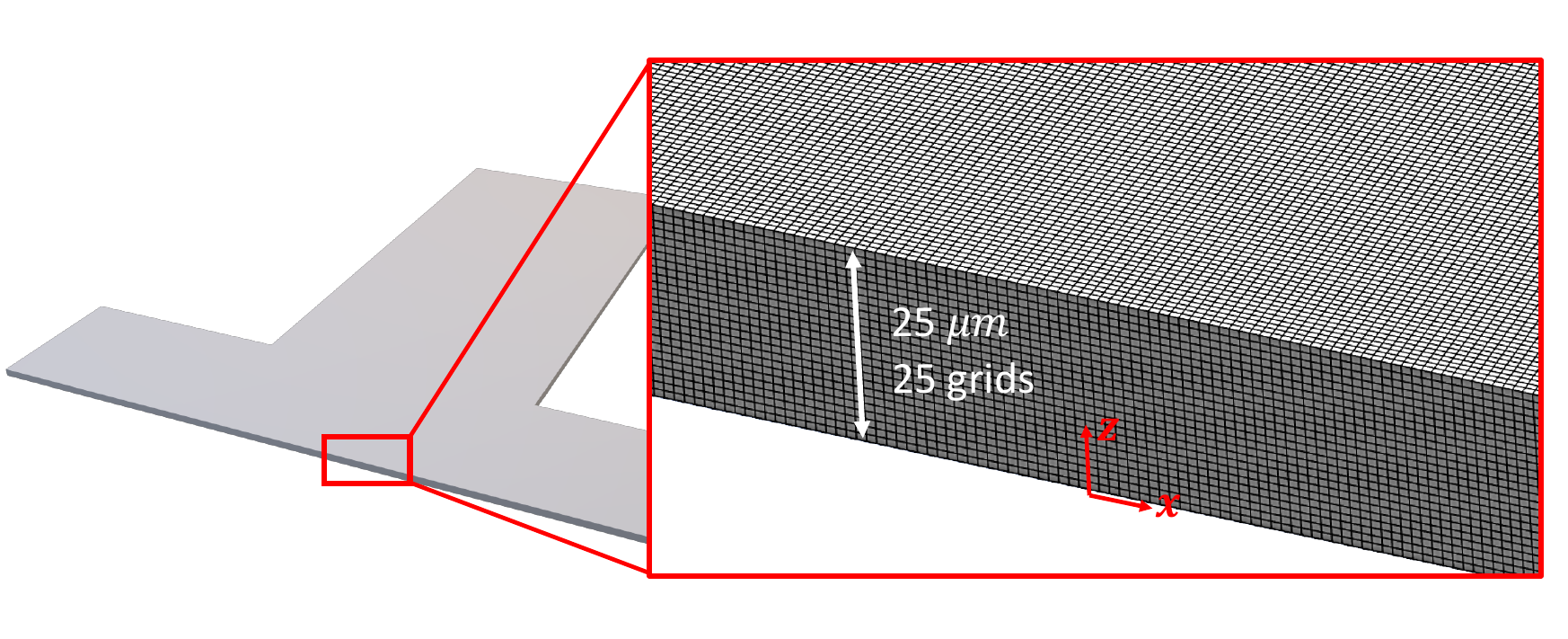}
    \caption{The zoomed-in picture of the mesh used in our simulations. It shows half of the domain (from $z=0$ to $H/2$). The channel gap is resolved by uniform regular mesh grids of size 1${\mu}m$.}
    \label{figs2}
    \end{figure}

To accurately capture the impact of three-dimensionality while minimizing numerical diffusion, we constructed a uniform, regular mesh across the entire T-junction domain, with 25 mesh grids across half the channel height ($z$ direction). We assume the flow to be symmetric with respect to the mid-plane of the microfluidic channel. The uniform cubic-shape mesh size is 1${\mu}m$; the total mesh number is around 83 million (Fig.~\ref{figs2}). The simulations are run on 200 cores on Yale high performance computing cluster.

\subsection{4. Diffusiophoretic and diffuisioosmotic mobilities}

In a binary Z-Z electrolyte and in the limit of thin Debye layer, the diffusiophoretic mobility of colloids can be written as \citep{prieve1984}:

    \begin{equation}
        \Gamma_{p} = \frac{\epsilon}{\eta} \left( \frac{k_BT}{Ze} \right)^2
        \left( \underbrace{\beta\frac{Ze\zeta_{p}}{k_BT}}_{\text{electrophoresis}} + \underbrace{4\text{lncosh} \left( \frac{Ze\zeta_{p}}{4k_BT} \right)}_{\text{chemiphoresis}} \right),
    \label{eqn.1}
    \end{equation}
where $\epsilon$ is the medium permittivity, $k_B$ is the Boltzmann constant, $Z$ is the valence of the electrolyte, $e$ is the elementary charge, $\eta$ is the fluid viscosity, $T$ is the temperature, $\zeta_p$ is the zeta potential of the colloid, and $\beta = \left(D_+ - D_-\right) / \left(D_+ + D_-\right)$ represents the difference in the mobility of cations and anions. The diffusioosmotic mobility $\Gamma_w$ of walls  can also be evaluated using this equation, replacing the zeta potential of particle $\zeta_p$ by that of the wall $\zeta_w$, and placing a minus sign in front of the expression to describes the fluid motion \citep{prieve1984}. In the normalization used in Fig.4 of the main text, we evaluated $\Gamma_p$ and $\Gamma_w$ using Eq.~\ref{eqn.1} for different salts assuming $\zeta_p=-70\text{mV}$, leading to $\Gamma_p=800 {\mu}m^2/s$, and $\zeta_w=-95, -105, -115\text{mV}$ for KCl, NaCl, and LiCl respectively \citep{kirby2004zeta1,Shin2016}.

\subsubsection{{4.1. Constant versus variable zeta potential models}}

The zeta potential of the colloids/surfaces are often described using either the i) constant zeta potential, or ii) constant charge approximations \citep{lee2023role, gupta2020}. More complex charge regulation models can also be utilized; however, they involve more parameters \citep{Markovich_2016, KROZEL1992365}. \\

\noindent \textbf{The constant zeta potential approximation} has been widely utilized and has been successful in reproducing many of the experimental observations involving solute-mediated transport of colloids \citep{abecassis2008, Shin2016}. However, this model has been shown to be quantitatively inaccurate for low/high solute concentrations, leading to over/under prediction of mobilities \citep{lee2023role, migacz2023convection}. In the presence of acid-base reactions, the zeta potential has been shown to change sign depending on the pH of the solution \citep{Suin2022}. In the limit of low solute concentrations, the thin Debye layer thickness approximation is also violated, rendering the expression above for the diffusiophoretic mobility inaccurate \citep{prieve1984,Shin2016, gupta2020}. \\

\noindent \textbf{The constant charge approximation} leads to a zeta potential that is a function of the bulk solute concentrations. In the case of binary electrolytes, the zeta potential has been described as
    \begin{equation}
        \zeta = a_0 + a_1 \text{log}(c),
    \label{eqn.5}
    \end{equation}
where $a_0$ and $a_1$ are material constants; for silica surface, $a_1$ is a function of cation types, i.e. $\text{Li}^+$, $\text{Na}^+$, and $\text{K}^+$, since they absorb on the surface to varying degrees \citep{kirby2004zeta1, kirby2004zeta2}. 

    \begin{figure}[b!]
    \centering
    \includegraphics[page=4, trim=80mm 5mm 80mm 5mm, clip, width=0.7\columnwidth]{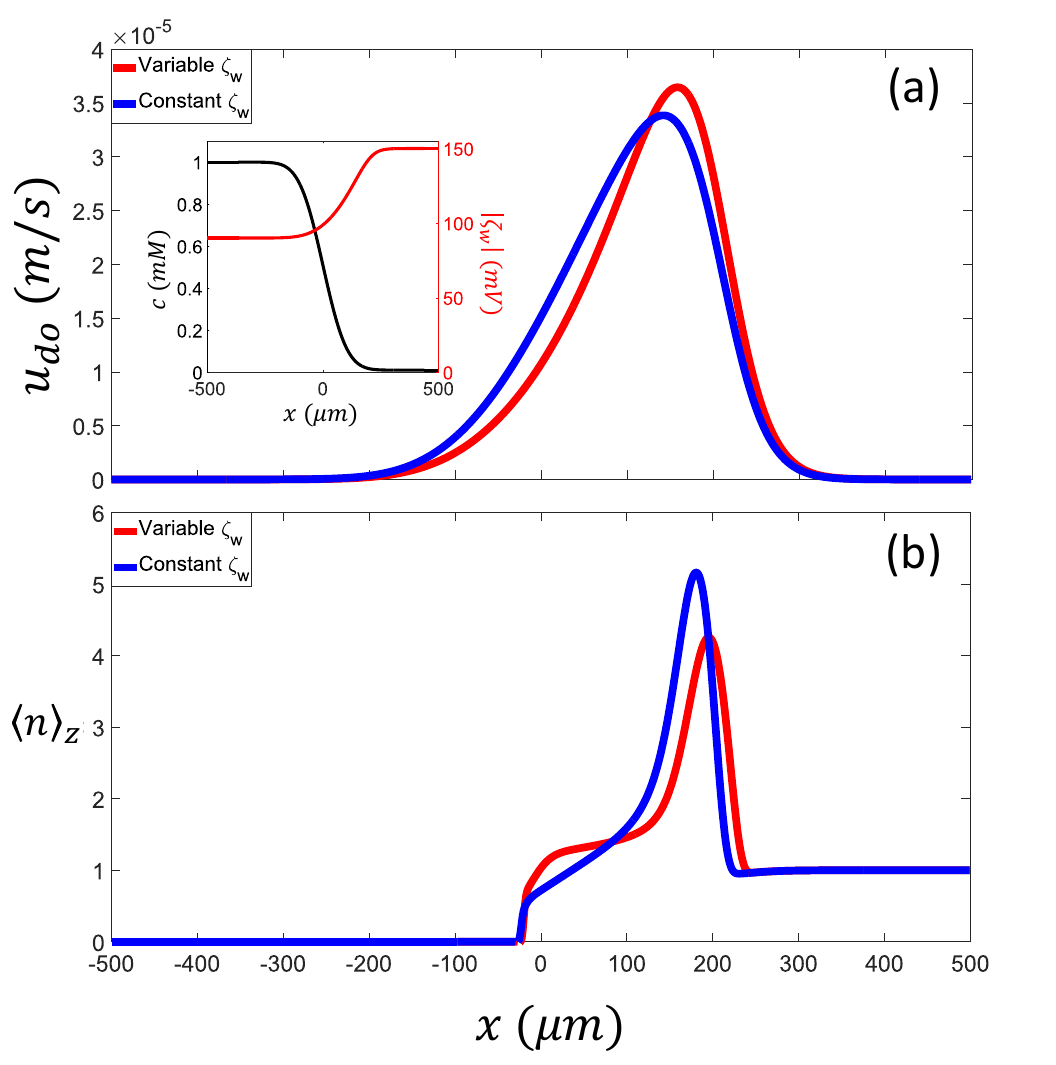}
    \caption{The comparison between predictions of the models with constant and variable zeta potential. (a) Diffusioosmotic velocity $u_{do}$ along x axis, at y=250$\mu m$. Inset: plot of solute concentration $c$ (left y axis) and the corresponding change in zeta potential magnitude $|\zeta_w|$ (right y axis) at channel wall. (b) Particle density $n$ (averaged along channel height) at y=250$\mu m$.}
    \label{figs3}
    \end{figure}

To understand how the use of the variable versus constant zeta potential models impacts the results of our numerical simulations, we conducted a simulation with the variable wall zeta potential model of Eq.~\ref{eqn.5} with $a_0\approx0$ and $a_1\approx30 mV$ \citep{kirby2004zeta1,chakra2023continuous}) (Fig.\ref{figs3} (a)). Using the representative values $c_L=1\text{mM}$, $c_R=0.01\text{mM}$, and $\text{Pe}\approx8$, we observed that the predicted diffusioosmotic velocity and the particle density in the two models is very close (Fig.\ref{figs3} (a,b)). Therefore, for simplicity, and to highlight the influence of competition between diffusioosmosis and diffusiophoresis on the colloidal trajectories, we use the constant zeta potential model in this work. It would certainly be interesting to explore the role of variable zeta potential in more details in the future.

\subsubsection{{4.2. The strength of diffusioosmotic mobility}}

\begin{figure*}[h!]
    \centering
    \includegraphics[page=5, trim=5mm 20mm 2mm 20mm, clip, width=1.0\textwidth]{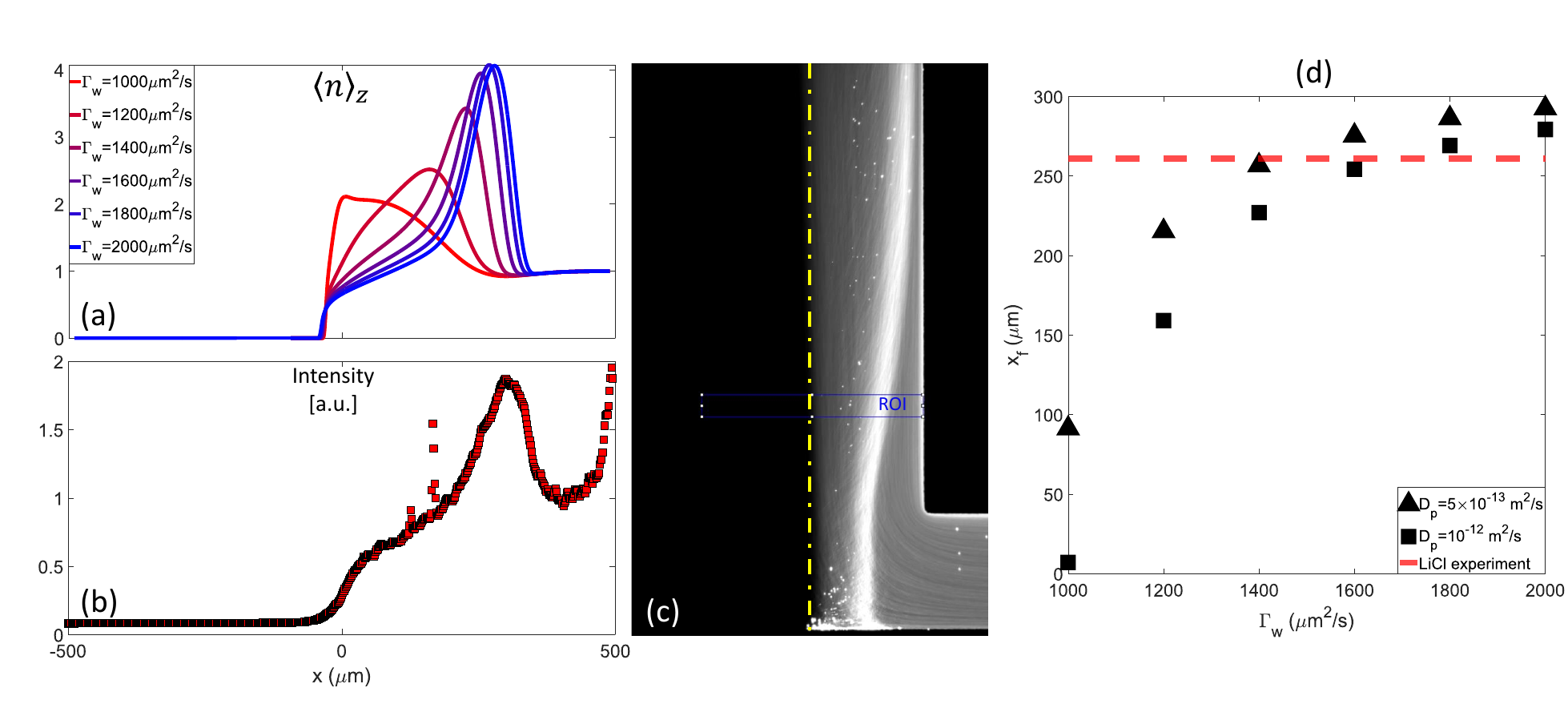}
    \caption{The calibration of diffusioosmotic mobility used in the numerical simulations. (a) The numerical simulation result of $z$-averaged particle intensity profile with different $\Gamma_w$ values. (b) The light intensity profile obtained from the region of interest (ROI) shown in the experimental image (c), which is the average of 250 long-exposure images. (d) The colloidal focusing position $x_f$ obtained from the numerical simulations versus $\Gamma_w$, with two different particle diffusivities $D_p$. Horizontal red dotted line shows the value of LiCl experiment with $\text{Pe}\approx8$ and $c_L/c_R=1000$.}
    \label{figs4}
    \end{figure*}

Measuring the zeta potential of surfaces is a complex task, often achieved indirectly from the streaming current characterization \citep{kirby2004zeta1}. In our numerical simulations, we treat the diffusioosmotic mobility of the wall as a fitting parameter, adjusting its value so that the colloid focusing location matches that of the experiments (Fig.~\ref{figs4}). 

The value of $x_f$ is taken from the peak location of the particle intensity profile  (Fig.~\ref{figs4}(a,b)). Experiment with LiCl, at $\text{Pe}\approx8$ and $c_L/c_R=1000$ is shown in Fig.~\ref{figs4}(c). Blue box shows the region of interest, i.e. rectangle at $y=2W$ with width $100{\mu}m$. We take the average of the intensity over the width and plot it over $x$ to get the profile in Fig.~\ref{figs4}(b). We find that $\Gamma_w \approx 1600{\mu}m^2/s$ leads to a close match to our experimental measurements (Fig.~\ref{figs4}(d)).

We use 1 micron size colloids with diffusivity $D_p\approx4.4\times10^{-13} m^2/s$ (by Stokes-Eintsein relation). This low diffusivity leads to a very large particle Pe number, increasing the computational cost. To circumvent this problem, we artificially increased the particle diffusivity in our simulations. We then checked to ensure this increase in particle diffusivity has minimal effect on the focusing location (Fig.~\ref{figs4}(d)). Therefore, to decrease the computational cost, we pick a slight enhanced value of particle diffusivity in our simulations ($D_p=1\times10^{-12}m^2/s$). 

\subsection{6. Self-similarity of the solute and colloid fields in the near-wall region}

In this section, we demonstrate that the partial differential equations governing the solute and colloidal fields can be simplified in the near-wall region using a similarity ansatz, simplifying the governing PDEs from three variables $(x,y,z)$ to two similarity variables $(\zeta, \eta)$. 

We consider a parallel flow with $y$ axis pointing in the flow direction, $z$ axis perpendicular to the top/bottom walls of the channels, measuring the distance from the mid-plane, and $x$ axis perpendicular to the flow direction. In the absence of solute gradients, the flow field can then be described as $(u_x,u_y,u_z) = (0,(3/2) U \left(1-(2 z/H)^2\right),0)$, i.e., a parabolic Poiseuille flow, where in the mid-plane $u_y = (3/2) U$ and at the walls, $z=\pm H/2$, the no-slip boundary condition is satisfied with $U$ representing the mean flow velocity.\\ 

\noindent \textbf{Solute field}\\

\noindent \ul{In the mid-plane}, $z=0$, the solute is advected along the $y$ direction with the flow velocity $(3/2)U$ and diffuses in the $x$ direction. This balance at steady state leads to
\begin{equation}
(3/2) U \frac{\partial c}{\partial y} = D_s \frac{\partial^2 c}{\partial x^2},
\end{equation}
which can be simplified using the similarity ansatz $\xi = x / \sqrt{D_s y/U}$, and $F(\xi) \equiv c(x,y)$ to
\begin{equation} 
- \frac{3}{4} \xi \frac{\partial F}{\partial \xi} = \frac{\partial^2 F}{\partial \xi^2},
\end{equation}
converting the original PDE in two variables to an ODE with one variable. \\

\noindent  \ul{Near the wall}, however, the fluid experiences a shear flow, $u_y \approx \dot{\gamma} \hat{z}$, where $\dot{\gamma} \sim U / H$ is the shear rate and $\hat{z} = z + H/2$ is the distance from the lower wall. The governing equation for the solute field then becomes
\begin{equation}
\dot{\gamma} \hat{z} \frac{\partial c}{\partial y} = D_s \left(\frac{\partial^2 c}{\partial x^2} + \frac{\partial^2 c}{\partial \hat{z}^2}\right),
\end{equation}
which can be simplified using the similarity ansatz $\zeta = \hat{z} / \left(D_s y/\dot{\gamma}\right)^{1/3}$, $\eta = x / \left(D_s y/\dot{\gamma}\right)^{1/3}$ and $F(\zeta,\eta) \equiv c(x,y,z)$ to
\begin{equation} 
- \frac{1}{3} \zeta^2 \frac{\partial F}{\partial \zeta} - \frac{1}{3} \zeta \eta \frac{\partial F}{\partial \eta} = \frac{\partial^2 F}{\partial \zeta^2} + \frac{\partial^2 F}{\partial \eta^2},
\end{equation}
converting the original PDE in three variables to one with two similarity variables. The scalings of these similarity variables indicate that the width of the solute diffusion zone near the wall scales as $\delta_s/H \sim \left(D_s y / U H^2\right)^{1/3} \sim \left(y/H\right)^{1/3} \textrm{Pe}^{-1/3}$, where $\textrm{Pe} = UH/D_s$ \citep{Ismagilov2000,salmon2007}. \\

\noindent \textbf{Colloid field}\\
In the microfluidic T-junction setup, both the solute and colloid fields are at a steady state. Neglecting the particle diffusivity, which is typically smaller by 3-4 orders of magnitude than the solute diffusivity, the following equation governs the colloid field:
\begin{equation}
\nabla \cdot \left( \left( \mathbf{u} + \mathbf{u}_{dp} \right) n \right) = 0,
\end{equation}
where $\mathbf{u}$ represents the fluid velocity and $\mathbf{u}_{dp} = \Gamma_p \nabla \ln c$ represents the diffusiophoretic velocity. Near the wall, we can approximate the the flow velocity as $\mathbf{u} = \left( -\Gamma_w \frac{\partial \ln c}{\partial x}, \dot{\gamma} \hat{z}, u_z \right)$, where the z-component of the flow velocity can be determined using the incompressibility constraint $\nabla \cdot \mathbf{u}=0$, leading to $u_z = \int_0^z \Gamma_w \frac{\partial^2 \ln c}{\partial x^2}~dz'$. Near the wall, the governing equation for the colloid field can then be written as
\begin{equation}
\Gamma_p n \left( \frac{\partial^2 \ln c}{\partial x^2} + \frac{\partial^2 \ln c}{\partial \hat{z}^2} \right) + \left(\Gamma_p-\Gamma_w \right) \frac{\partial n}{\partial x} \frac{\partial \ln c}{\partial x} + \dot{\gamma} \hat{z} \frac{\partial n}{\partial y} + \left( \Gamma_p \frac{\partial \ln c}{\partial \hat{z}} + \Gamma_w \int_0^{\hat{z}} \frac{\partial^2 \ln c}{\partial x^2}~dz' \right) \frac{\partial n}{\partial \hat{z}} = 0.
\end{equation}

We postulate that the above partial differential equation can be simplified using the following similarity ansatz: $F(\zeta,\eta) \equiv c(x,y,z)$, $G(\zeta,\eta) \equiv n(x,y,z)$ with $\zeta = \hat{z} / \left(D_s y/\dot{\gamma}\right)^{1/3}$, $\eta = x / \left(D_s y/\dot{\gamma}\right)^{1/3}$, i.e., the same similarity variables governing the solute field. Substituting these similarity variables into the governing equation for the colloid field, we arrive at the following:
\begin{align}
& \left( \frac{\partial^2 \ln F}{\partial \zeta^2} + \frac{\partial^2 \ln F}{\partial \eta^2} \right) 
+ \left(1 - \frac{\Gamma_w}{\Gamma_p} \right) \frac{\partial \ln F}{\partial \eta} \frac{\partial \ln G}{\partial \eta} 
- \frac{D_s}{3 \Gamma_p} \left( \zeta^2 \frac{\partial \ln G}{\partial \zeta} + \zeta \eta \frac{\partial \ln G}{\partial \eta} \right)  \nonumber \\ 
& +  \frac{\partial \ln F}{\partial \zeta} \frac{\partial \ln G}{\partial \zeta} 
+ \frac{\Gamma_w}{\Gamma_p} \zeta \frac{\partial \ln G}{\partial \zeta} \frac{\partial^2 \ln F}{\partial \eta^2} = 0,
\end{align}
which is a partial differential equation for the colloid field with two similarity variables $\zeta$ and $\eta$. Unlike the governing equation for the solute field, here the coefficients involve the solute diffusivity, and osmotic and phoretic mobilities. The success of the similarity ansatz in reducing the governing equation from three to two variables, indicates that the relevant length-scale for the colloid field near the wall is the solute diffusion width as also arrived at via scaling arguments in the main text.

We can further simplify this equation near the wall, taking $\zeta \to 0$:
\begin{equation}
\underbrace{\left( \frac{\partial^2 \ln F}{\partial \zeta^2} + \frac{\partial^2 \ln F}{\partial \eta^2} \right)}_{\text{compressibility } (\nabla \cdot \mathbf{u}_{dp})} 
+ \underbrace{\left(1 - \frac{\Gamma_w}{\Gamma_p} \right) \frac{\partial \ln F}{\partial \eta} \frac{\partial \ln G}{\partial \eta}}_{\text{diffusiophoresis vs diffusioosmosis along the wall}} 
+  \underbrace{\frac{\partial \ln F}{\partial \zeta} \frac{\partial \ln G}{\partial \zeta}}_{\text{diffusiophoresis perpendicular to the wall}}  = 0,
\end{equation}
which can be further simplified by keeping only the derivatives along the wall to 
\begin{equation}
\frac{\partial^2 \ln F}{\partial \eta^2} 
+ {\left(1 - \frac{\Gamma_w}{\Gamma_p} \right) \frac{\partial \ln F}{\partial \eta} \frac{\partial \ln G}{\partial \eta}} = 0,
\end{equation}
which is the same as Eq. 7 in the main text.  

\bibliography{SM}